\documentclass{article}
\usepackage{amsmath,amssymb, amsthm,amstext}
\usepackage{graphicx}
\usepackage{color}
\usepackage{array, enumerate}
\usepackage{bm}
\usepackage{multirow}
\usepackage{braket}
\usepackage{txfonts}
\usepackage{booktabs}
\usepackage{afterpage}
\usepackage{subfigure}
\usepackage{rotating} 
\usepackage{float}

\usepackage{ulem}
\usepackage{xcolor}
\newcommand{\newadded}[1]{{\textcolor{red}{#1}}}
\newcommand{\old}[1]{{\color{red}{{}}}}

\usepackage{geometry}
\usepackage{indentfirst}

\title{Bilateral constraints on proton Lorentz violation effects\thanks{P.~He \& B.-Q.~Ma, \href{https://doi.org/10.3847/1538-4357/adad6c}{Astrophys.~J. 981 (2025) 45.}}}

\author{
    Ping He\\
    School of Physics, Peking University, Beijing 100871,China\\
    \and
    Bo-Qiang Ma\\
    School of Physics, Peking University, Beijing 100871,China\\
    Center for High Energy Physics, Peking University, Beijing 100871, China\\
    School of Physics, Zhengzhou University, Zhengzhou 450001, China,
    \texttt{mabq@pku.edu.cn}
}

\date{Received 2024 September 15; revised 2025 January 21; accepted 2025 January 21; published 2025 February 25}

\begin{document}

\maketitle

\begin{abstract}
    Since the early reports of events beyond the Greisen-Zatsepin-Kuzmin (GZK) cutoff, the investigation of ultrahigh-energy cosmic rays has emerged as a fundamental method for testing Lorentz Invariance violation (LV) effects. Recent advances in observational capabilities have resulted in more stringent constraints on LV parameters. This study delves into the potentially anomalous phenomena arising from subluminal and superluminal LV effects, encompassing aspects such as proton decay and atypical threshold behavior of photo-pion production from protons within the GZK region. High-energy proton observations in cosmic rays have imposed stringent constraints on superluminal proton LV effects, while the confirmation of the GZK cutoff has established a robust boundary for the anomalous phenomena associated with subluminal proton LV effects. The research provides rigorously bilateral constraints on proton LV effects from both subluminal and superluminal standpoints.
\end{abstract}

\noindent
\textbf{keywords:} Lorentz invariance violation, ultrahigh-energy cosmic rays, Greisen-Zatsepin-Kuzmin (GZK) cutoff,  subluminal Lorentz violation,  superluminal Lorentz violation

\section{Introduction} 

The Greisen-Zatsepin-Kuzmin (GZK) cutoff, predicted in 1966, posits that ultrahigh-energy cosmic rays~(UHECRs, with energies above $5\times10^{19}~\mathrm{eV}$) interact with cosmic microwave background~(CMB) radiation, resulting in energy loss and a distinct decline in the cosmic-ray spectrum~\cite{P1-Greisen-1966-end,P2-Zatsepin-1966-upper}. Early cosmic-ray experiments, including Akeno, Haverah Park, Yakutsk, and Fly's Eye, initially encountered challenges in confirming the GZK cutoff~(for a comprehensive review, refer to Ref.~\cite{P3-Sokolsky-2007-Highest}). Some experiments documented cosmic rays with energies surpassing the GZK threshold~\cite{P4-Bird-1994-detection,P5-Takeda-1998-extension}, giving rise to the ``GZK paradox:" the absence of nearby astrophysical sources capable of explaining these high-energy occurrences. The potential interpretation of these post-GZK events as indicators of Lorentz invariance violation~(LV) has been investigated~\cite{P20-Coleman-1998-high,P21-Coleman-1998-evading,P22-Amelino-2000-planck,P23-Amelino-2001-space,P25-Gonzalez-1997-vacuum,P26-Sato-2000-extremely,P27-Bertolami-1999-proposed,P28-Aloisio-2000-probing}. In scenarios where Lorentz invariance, a fundamental tenet of relativity, is marginally disrupted, UHECRs may circumvent anticipated energy depletion from interactions with the CMB, potentially elucidating the observations of cosmic rays beyond the GZK cutoff.\\

Cosmic rays, particularly UHECRs, present a distinct advantage in the investigation of LV.
Cosmic rays are high-energy particles that originate from outer space, primarily consisting of protons, with a smaller fraction made up of heavier nuclei (such as helium nuclei or alpha particles), as well as electrons, gamma-rays, and neutrinos (for comprehensive reviews on this topic, see Refs.~\cite{P86-Berezinsky-1990-Astrophysics,P87-Anastasi-2024-The}).
UHECRs possess exceptionally high energies that surpass those attainable in human-made particle accelerators~\footnote{The highest energy value of human-made particles on Earth today is about $10^{13}~\mathrm{eV}$, produced by the Large Hadron Collider, and the highest-energy particles ever observed in cosmic rays is above $10^{20}~\mathrm{eV}$~\cite{P88-TelescopeArray-2023-An}. 
This value is equivalent to the energy of $4\times 10^5~\mathrm{GeV}$, which is $400~\mathrm{TeV}$ in the equivalent center-of-mass~(c.m.) frame of the $p+p$ collision~(see Fig.~1 in ref.~\cite{Pierog:2013mbl}).}. 
Traveling across vast cosmological distances, these particles can accumulate even tiny LV effects over their journey, potentially leading to detectable outcomes~\cite{P19-H18-He-2022-lorentz}. 
Moreover, the diverse array of particles within cosmic rays, encompassing protons, electrons, and photons, offers a diverse experimental landscape for exploring various facets of LV~(for comprehensive review, refer to Refs.~\cite{P103-Kostelecky-2008-date,P104-Addazi-2021-quamtum,P105-AlvesBatista-2023-white}).\\

The GZK ``paradox," which refers to the observation of cosmic-ray events with energies exceeding the GZK threshold in the 1990s~\cite{P4-Bird-1994-detection,P5-Takeda-1998-extension}, marked a significant turning point in the investigation of LV.
This apparent ``paradox" prompted a flurry of exotic theoretical proposals aimed at explaining the phenomenon,~\footnote{Some hypotheses suggested the existence of new components of cosmic rays, such as magnetic monopoles~\cite{P13-Kephart-1995-magnrtic}, ``Z-boson bursts"~\cite{P14-Weiler-1997-cosmic}, and decay products of hypothetical superheavy relic particles~\cite{P15-Berezinsky-1997-ultrahigh}. Other theories posited that these cosmic rays above the GZK cutoff might originate from unidentified sources. Additionally, some explanations attributed these post-GZK events to propagation effects, suggesting that cosmic rays experience significant deflections due to the structure of extragalactic magnetic fields~\cite{P12-Farrar-1999-GZK}.} igniting discussions on LV within the broader context of fundamental physics~\cite{P20-Coleman-1998-high,P21-Coleman-1998-evading,P22-Amelino-2000-planck,P23-Amelino-2001-space,P25-Gonzalez-1997-vacuum,P26-Sato-2000-extremely,P27-Bertolami-1999-proposed,P28-Aloisio-2000-probing}.
In scenarios where Lorentz invariance is only marginally disrupted, UHECRs may circumvent the expected energy depletion from interactions with the CMB, potentially providing an explanation for the observed cosmic rays that exceed the GZK cutoff.
Coleman and Glashow proposed a perturbative framework in which small noninvariant terms were introduced into the standard model Lagrangian. This approach allows for potential violations of strict Lorentz invariance that could suppress or eliminate inelastic collisions between cosmic-ray nuclei and CMB photons, thereby relaxing or removing the GZK cutoff~\cite{P20-Coleman-1998-high,P21-Coleman-1998-evading}.
Amelino-Camelia developed a general phenomenological framework for describing the deformation of Lorentz invariance, suggesting that LV effects could emerge from the nontrivial short-distance structure of spacetime~\cite{P22-Amelino-2000-planck,P23-Amelino-2001-space}.
Jacobson, Liberati, and Mattingly utilized various astrophysical phenomena to derive limits on LV parameters, providing insights into potential quantum gravity effects, with a particular emphasis on quantum electrodynamics~(QED) phenomena, including photon decay, the vacuum Cherenkov effect, and synchrotron radiation~\cite{P18-H6-Mattingly-2002-threshold,P59-Jacobson-2002-Threshold,P89-Jacobson-2002-A,P90-Jacobson-2003-New,P91-Jacobson-2004-Quantum,P92-Jacobson-2004-Astrophysical}.
Galavani and Sigl investigated the compositional changes of photons in ultrahigh-energy spectra under LV effects, demonstrating that the propagation of photons, electrons, and positrons above \(\sim10^{19}~\mathrm{eV}\) can be significantly altered by LV. They extended the constraints on LV effects for photons and electrons to both first and second orders~\cite{P93-Galaverni-2007-Lorentz,P94-Galaverni-2008-Lorentz}.
Martínez-Huerta and Pérez-Lorenzana employed the dispersion relation under LV effects to calculate the squared probability amplitude for vacuum Cherenkov radiation and photon decay, deriving general rates for these processes at leading order for any order $n$~\cite{P95-Martinez-Huerta-2016-Vacuum,P96-Martinez-Huerta-2017-Effects,P97-Martinez-Huerta-2017-Restrictions,P98-Martinez-Huerta-2017-Photon}.\\

The exploration of LV through cosmic rays has made significant strides in recent times, propelled by enhanced observational technologies and refined theoretical frameworks. Ongoing research continues to exploit cosmic-ray observations to delve into LV, focusing on phenomena such as photon dispersion relations, particle decay mechanisms, and the behavior of cosmic rays at ultrahigh energies~(for an overview, refer to Ref.~\cite{P19-H18-He-2022-lorentz}). Notably, constraints derived from the Large High Altitude Air Shower Observatory (LHAASO) represent some of the most stringent constraints on photon and electron LV effects to date, markedly narrowing the parameter space for theoretical models integrating LV effects~\cite{P80-H3-lhaaso-2021-exploring,P42-Li-2021-ultrahigh,P81-H5-chen-2021-strong,P82-H7-li-2022-testing,P83-H43-He-2022-joint,P84-He-2023-Comprehensive}.\\

The existence of the GZK cutoff presents an opportunity to constrain LV effects. In 2004, the High Resolution Fly’s Eye (HiRes) experiment provided data clearly indicating a termination in the cosmic-ray flux, aligning with the predicted GZK cutoff~\cite{P7-Thomson-2004-new,P8-HiRes-2004-measyrement,P9-HiRes-2005-observation,P10-HiRes-2008-first}. As more observational instruments have been deployed, the presence of the GZK cutoff has been further validated~\cite{P29-PierreAuger-2020-features,P30-PierreAuger-2020-measurement,P35-Abbasi-2023-the}. The observation of the GZK cutoff also imposes stringent constraints on LV parameters of protons from theoretical analyses~\cite{P38-Xiao-2008-lorentz,P39-Bi-2008-testing,P40-Stecker-2009-searching}; for instance, the Pierre Auger Observatory obtains constraints on LV in both the electromagnetic and hadronic sectors by comparing observational data with LV simulations of the energy spectrum, cosmic-ray composition, and upper limits on photon flux~\cite{P55-Lang-2020-Testing, P57-PierreAuger-2021-testing}.
Despite this, there remain ultrahigh-energy events that present intriguing possibilities, such as the Telescope Array experiment's detection of an exceptionally energetic particle registered at $2.44\pm0.29~(\rm{stats.})^{+0.51}_{-0.71}~(\rm{syst.})\times10^{20}~\mathrm{eV}$~\cite{P88-TelescopeArray-2023-An}, potentially indicating gaps in our understanding of particle physics.\\

The preceding constraints stemming from the GZK structure are examined in the context of UHECRs interacting with CMB radiation, representing constraints specifically on subluminal proton LV effects. The aim of our investigation is to explore the impact of proton LV effects from both subluminal and superluminal standpoints. We delve into the potential anomalous phenomena arising from subluminal and superluminal LV effects, including proton decay and the atypical threshold behaviors observed in the interaction of UHECRs with CMB radiation. Our analysis yields constraints on proton LV effects from both subluminal and superluminal perspectives, thus establishing bilateral constraints on proton LV effects.\\

\section{Phenomenological Model and Analysis}

We adopt a model-independent approach to introduce the LV for a particle with mass $m$ by modifying the dispersion relation
\begin{eqnarray}\label{modification dispersion relation}
   E^2 &=& F(E, p; m, \eta_\mathrm{n})  \notag \\ 
   &=& p^2+m^2-\eta_\mathrm{n}p^2\left(\frac{p}{E_\mathrm{Pl}}\right)^n,
\end{eqnarray}
where $E_\mathrm{Pl}\simeq1.22\times10^{19}~\mathrm{GeV}$ is the Plank energy and $\eta_\mathrm{n}$ is the $n$th-order LV parameter. 
Since the LV effects are very tiny, we generally expect that LV effects could have observable effects only in the extremely high-energy region, and the LV modifications would be suppressed by the Planck energy. 
We can find the corresponding velocity $v(E)=\frac{\partial E}{\partial p}\approx 1-\frac{m^2}{2p^2}-\eta_\mathrm{n}\frac{n+1}{2}(\frac{E}{E_\mathrm{Pl}})^n$. 
For $\eta_\mathrm{n}>0$, the greater the energy, the slower the velocity, corresponding to the subluminal LV effect.
For $\eta_\mathrm{n}<0$, the greater the energy, the faster the velocity, corresponding to the superluminal LV effect.
For different particles, there are corresponding dispersion relations. 
In this article, we care about the dispersion relations of protons, pions and photons
\begin{equation}\label{proton photon and pion modification dispersion relation}
    \begin{cases}
    E^2_\mathrm{p} = p^2_\mathrm{p}+m^2_\mathrm{p}-\eta_\mathrm{p,n}p^2_\mathrm{p}\left(\frac{p_\mathrm{p}}{E_\mathrm{Pl}}\right)^n,  &  (1*) \mathrm{\quad for \quad proton};\\
    {\omega}^2 = k^2-\xi_\mathrm{n} k^2\left(\frac{k}{E_\mathrm{Pl}}\right)^n,  &  (2*) \mathrm{\quad for \quad photon};\\
    E^2_\pi = p^2_\pi+m^2_\pi-\eta_\mathrm{\pi,n}p^2_\pi\left(\frac{p_\pi}{E_\mathrm{Pl}}\right)^n,  &  (3*) \mathrm{\quad for \quad pion},\\
    \end{cases}
\end{equation}
For convenience, if we only consider the linear modification with $n=1$, we set $\eta_\mathrm{p, 1}\equiv\eta_\mathrm{p}$, $\xi_\mathrm{1}\equiv\xi$, and $\eta_\mathrm{\pi, 1}\equiv\eta_\pi$.
Currently, there are many theories that can induce a violation of Lorentz symmetry~\cite{P19-H18-He-2022-lorentz}, and these theories may cause ambiguities when studying LV effects. 
By choosing this phenomenological framework, we can study the proton reaction with photons under LV effects model independently.
In the following, we discuss the potentially abnormal phenomena: the proton decay and the abnormal threshold behavior of the reaction of UHECRs interacting with CMB radiation~\cite{P85-He-2024-Abnormal}.
We discuss the proton LV effect from both the subluminal and superluminal sides. 
Then, we get the corresponding constraints on LV parameters by confronting with current observations.\\

\subsection{Proton decay $\mathrm{p}\to \mathrm{p}+\gamma$}

A high-energy proton with momentum $p$ decays into a proton with momentum $yp$~($y\in[0,1]$) and a photon with momentum $(1-y)p$.
The energy-momentum conservation relation is
\begin{equation}
\label{energy momentum conservation of proton dacay}
 E_\mathrm{proton}(p)=E_\mathrm{proton}(yp)+\omega_\mathrm{photon}[(1-y)p],
\end{equation}
\newadded{where $E_\mathrm{proton}(yp)$ refers to the energy of the proton with momentum $yp$ and $\omega_\mathrm{photon}[(1-y)p]$ refers to the energy of the photon with momentum $(1-y)p$}.
We introduce the LV effect though the modification dispersion relation Eq.~(\ref{proton photon and pion modification dispersion relation}), and we get 
\begin{equation}\label{detail energy momentum conservation of proton dacay}
    p[1+\frac{m^2_\mathrm{p}}{2p^2}-\frac{\eta_\mathrm{p,n}}{2}(\frac{p}{E_\mathrm{Pl}})^n]=yp[1+\frac{m^2_\mathrm{p}}{2(yp)^2}-\frac{\eta_\mathrm{p,n}}{2}(\frac{yp}{E_\mathrm{Pl}})^n]+(1-y)p[1-\frac{\xi_\mathrm{n}}{2}(\frac{(1-y)p}{E_\mathrm{Pl}})^n].
\end{equation}
After algebraic operations, we get
\begin{equation}\label{energy and LV parameter of ptoton decay}
    \frac{m^2_\mathrm{p}E_\mathrm{Pl}^n}{p^{n+2}}=\frac{y-y^{n+2}}{y-1}\eta_\mathrm{p,n}+y(1-y)^n\xi_\mathrm{n}.
\end{equation}
Eq.~(\ref{energy and LV parameter of ptoton decay}) means that finding the proton decay threshold $p_\mathrm{th}^\mathrm{decay}$ is equivalent to finding the minimum value of $p$ on the left side of Eq.~(\ref{energy and LV parameter of ptoton decay}), and correspondingly, maximizing the right side of Eq.~(\ref{energy and LV parameter of ptoton decay}). 
The $y$-value satisfying Eq.~(\ref{energy and LV parameter of ptoton decay}) reflects the momentum distribution after proton decay, where $y=1$ (or $0$) means that the product proton~(or photon) obtains the most momentum and $y=1/2$ means that the momentum distribution is equal between the product proton and photon. 
When $\eta_\mathrm{p,n}\to0$ and $\xi_\mathrm{n}\to0$, the maximum value on the right side tends to zero, and $p_\mathrm{th}^\mathrm{decay}\to +\infty$.
This corresponds to the situation where protons cannot decay in the classical case as expected, since any situation must return to the classical case when the LV effects approach zero.\\

If we only consider the linear~($n=1$) modification, Eq.~(\ref{energy and LV parameter of ptoton decay}) becomes
\begin{equation}\label{linear energy and LV parameter of ptoton decay}
    \frac{m^2_\mathrm{p}E_\mathrm{Pl}}{p^3}=-y(y+1)\eta_\mathrm{p}+y(1-y)\xi.
\end{equation}
Next let us see the corresponding proton decay threshold in different LV modification parameter configurations, and how the observation sets constraints on LV modification parameters.
In three kinds of different LV parameter configuration, there are three thresholds of the proton decay.
\begin{enumerate}   
    \item	
    When $\eta_\mathrm{p}>0$ and $\xi-\eta_\mathrm{p}<0$, the right side of Eq.~(\ref{linear energy and LV parameter of ptoton decay}) is negative ($\le0$) within the range of values. 
    Thus there is no proton decay, i.e., the proton decay threshold is
    \begin{equation}
         p_\mathrm{th}^\mathrm{decay}=+\infty.
     \end{equation}        
     \item	
     When $\eta_\mathrm{p}<0$ and $\xi+3\eta_\mathrm{p}<0$, the maximum value on the right side of Eq.~(\ref{linear energy and LV parameter of ptoton decay}) is $-2\eta_\mathrm{p}$, which is taken at $y=1$.
     The proton decay threshold is
     \begin{equation}
         p_\mathrm{th}^\mathrm{decay}=(\frac{m^2_\mathrm{p}E_\mathrm{Pl}}{-2\eta_\mathrm{p}})^{1/3}.
     \end{equation}     
      \item	
      When $\xi-\eta_\mathrm{p}>0$ and $\xi+3\eta_\mathrm{p}>0$, the maximum value on the right side of Eq.~(\ref{linear energy and LV parameter of ptoton decay}) is $(\xi-\eta_\mathrm{p})^2/(4\xi+4\eta_\mathrm{p})$, which is taken at $y=(\xi-\eta_\mathrm{p})/(2\xi+2\eta_\mathrm{p})$. 
      The proton decay threshold is
     \begin{equation}
         p_\mathrm{th}^\mathrm{decay}=(\frac{4(\xi+\eta_\mathrm{p})m^2_\mathrm{p}E_\mathrm{Pl}}{(\xi-\eta_\mathrm{p})^2})^{1/3}.
     \end{equation}
\end{enumerate}
\begin{figure}[H]
    \centering
    \includegraphics[scale=0.98]{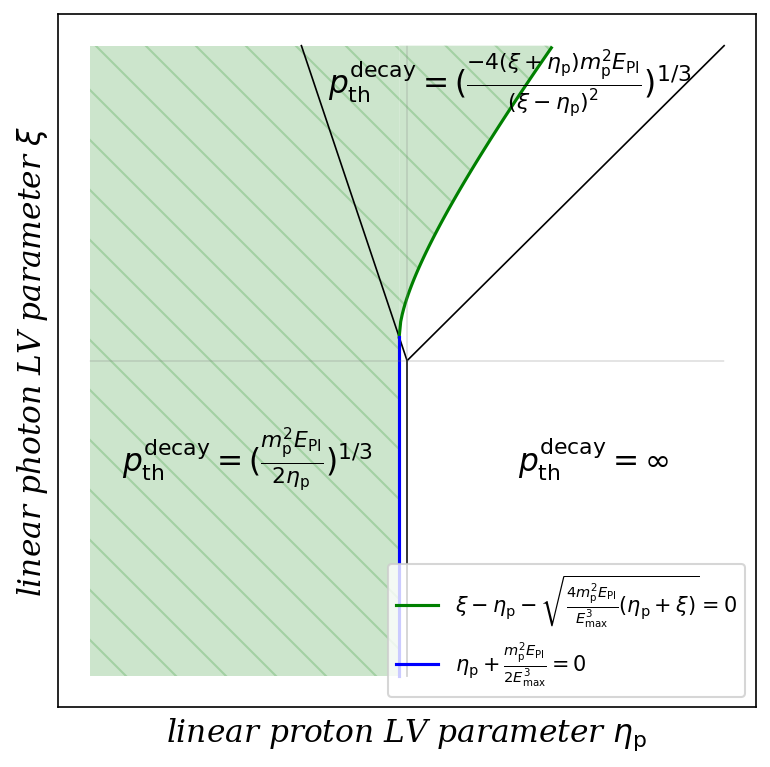}
    \caption{Proton decay constraint on the proton-photon LV parameter plane from the highest-energy proton ever detected.}
    \label{fig proton decay}
\end{figure}

When the proton energy exceeds these thresholds, proton decay can occur, which results in a rapid drop in the proton energy. 
On the other hand, observing a high-energy proton, whose energy is $E_\mathrm{max}$, means that $E_\mathrm{max}$ does not reach the decay threshold $p_\mathrm{th}^\mathrm{decay}$: $p_\mathrm{th}^\mathrm{decay}>E_\mathrm{max}$.
In different LV parameter configurations, the corresponding constraints on LV parameters are~(we also show these constraints in Fig.~\ref{fig proton decay})
\begin{equation*}
    \begin{cases}
    \mathrm{no\quad constraints}   &  (i)\quad\eta_\mathrm{p}>0\quad \mathrm{and}  \quad \xi-\eta_\mathrm{p}<0;\\
    \eta_\mathrm{p}>-\frac{m^2_\mathrm{p}E_\mathrm{Pl}}{2E_\mathrm{max}^3}  &  (j)\quad\eta_\mathrm{p}<0\quad \mathrm{and}  \quad \xi+3\eta_\mathrm{p}<0;\\
    0<\xi-\eta_\mathrm{p}<\sqrt{\frac{4m^2E_\mathrm{Pl}(\xi+\eta_\mathrm{p})}{E_\mathrm{max}^3}}  &   (k)\quad\xi-\eta_\mathrm{p}>0\quad \mathrm{and}  \quad \xi+3\eta_\mathrm{p}>0.\\
    \end{cases}
\end{equation*}
If we suppose that the $2.44~\mathrm{EeV}$ event from the Telescope Array experiment is the highest-energy proton, we can get~\cite{P88-TelescopeArray-2023-An}, the strict constraints on proton-photon LV parameter plane:
\begin{equation*}
    \begin{cases}
    \mathrm{no\quad constraints}   &  (i)\quad\eta_\mathrm{p}>0\quad \mathrm{and}  \quad \xi-\eta_\mathrm{p}<0;\\
    \eta_\mathrm{p}>-3.69\times10^{-16}  &  (j)\quad\eta_\mathrm{p}<0\quad \mathrm{and}  \quad \xi+3\eta_\mathrm{p}<0;\\
    0<\xi-\eta_\mathrm{p}<\sqrt{2.96\times10^{-15}(\xi+\eta_\mathrm{p})}  &   (k)\quad\xi-\eta_\mathrm{p}>0\quad \mathrm{and}  \quad \xi+3\eta_\mathrm{p}>0.\\
    \end{cases}
\end{equation*}\\

From Fig.~\ref{fig proton decay} we can see that the high-energy proton event from the cosmic-ray observation have set very strict constraints on the proton and photon LV parameters.
If we presuppose that there is no photon LV effect~($\xi=0$), the constraint on the proton LV parameter is $\eta_\mathrm{p}>-\frac{m^2_\mathrm{p}E_\mathrm{Pl}}{2E_\mathrm{max}^3}$, and it is a strict constraint on proton superluminal linear LV modification.
If we suppose that the $2.44~\mathrm{EeV}$ event from the Telescope Array experiment is the highest-energy proton we can get~\cite{P88-TelescopeArray-2023-An}, we get the strict superluminal constraint $\eta_\mathrm{p}>-3.69\times10^{-16}$.\\

\subsection{Proton-Photon reaction $\mathrm{p}+\gamma\to\mathrm{p}+\pi^0$}

A high-energy proton with momentum $p$ scatters with a photon with momentum $-k$ and produces a secondary proton with momentum $y(p-k)$~($y\in[0,1]$) and a pion with momentum $(1-y)(p-k)$. 
The energy-momentum conservation relation is
\begin{equation}\label{energy momentum conservation of proton scatter}
 E_\mathrm{proton}(p)+\omega_\mathrm{photon}(-k)=E_\mathrm{proton}[y(p-k)]+E_\mathrm{pion}[(1-y)(p-k)].  
\end{equation}
We introduce the LV effect though the modification dispersion relation Eq.~(\ref{proton photon and pion modification dispersion relation}), and we get 
\begin{equation}\label{detail energy momentum conservation of proton scatter}
   \begin{aligned}
    p[1+\frac{m^2_\mathrm{p}}{2p^2}-\frac{\eta_\mathrm{p,n}}{2}(\frac{p}{E_\mathrm{Pl}})^n]+k=y(p-k)[1+\frac{m^2_\mathrm{p}}{2y^2(p-k)^2}-\frac{\eta_\mathrm{p,n}}{2}\frac{y^n(p-k)^n}{E^n_\mathrm{Pl}}]\\+(1-y)(p-k)[1+\frac{m^2_\pi}{2(1-y)^2(p-k)^2}-\frac{\eta_\mathrm{\pi,n}}{2}\frac{(1-y)^n(p-k)^n}{E^n_\mathrm{Pl}}],
   \end{aligned}
\end{equation}
where LV parameters $\eta_\mathrm{p,n}$ and $\eta_\mathrm{\pi,n}$ and photon momentum $k$ are small quantities.
After algebraic operations, we get
\begin{equation}\label{energy and LV parameter of ptoton scatter}
    E^n_\mathrm{Pl}\{\frac{4k}{p^{n+1}}+\frac{1}{p^{n+2}}[\frac{-(1-y)^2m^2_\mathrm{p}-ym^2_\mathrm{\pi}}{y(1-y)}]\}=\eta_\mathrm{p,n}(1-y^{n+1})-\eta_\mathrm{\pi,n}(1-y)^{n+1}.   
\end{equation}
Similarly to the discussion of proton decay, the results of Eq.~(\ref{energy and LV parameter of ptoton scatter}) are the thresholds of the proton and photon scatter reactions.
When $\eta_\mathrm{p,n}\to0$ and $\eta_\mathrm{\pi,n}\to0$, Eq.~(\ref{energy and LV parameter of ptoton scatter}) returns to the classical case, and we obtain the classical proton scatter threshold $E_\mathrm{th}=\frac{1}{4\omega_\mathrm{photon}}[(m_\mathrm{p}+m_{\pi})^2-m_\mathrm{p}^2]$ in the energy-momentum distribution $y=\frac{m_\mathrm{p}}{m_\mathrm{p}+m_\pi}$.
For CMB photons, the target photon energies obey a thermal distribution with temperature $T=2.73~\mathrm{K}$, or $\omega_0\equiv kT=2.35\times10^{-4}~\mathrm{eV}$. 
If we simply consider the CMB characteristic energy as the photon reaction energy, the corresponding proton reaction threshold is $E_\mathrm{th}=2.88\times10^{20}~\mathrm{eV}$. 
Above this threshold, protons should experience extensive collisions to lose energy, resulting in a sharp decrease in the spectrum, i.e., the GZK mechanism~\cite{P1-Greisen-1966-end,P2-Zatsepin-1966-upper}.\\

Unlike proton decay, there are $p^{n+1}$ and $p^{n+2}$ in Eq.~(\ref{energy and LV parameter of ptoton scatter}), so it is difficult to get an analytical expression for the results.
We choose numerical calculation to obtain the thresholds of the proton and photon scatter reactions.
Before that, we get some properties of Eq.~(\ref{energy and LV parameter of ptoton scatter}), which helps us investigate the thresholds.
If we multiply Eq.~(\ref{energy and LV parameter of ptoton scatter}) by $p^{n+2}\frac{y(1-y)}{(1-y)^2m^2_\mathrm{p}+ym^2_\mathrm{\pi}}$, we get
\begin{equation}\label{A2}
    \frac{y(1-y)[\eta_\mathrm{\pi,n}(1-y)^{n+1}-\eta_\mathrm{p,n}(1-y^{n+1})]}{E^n_\mathrm{Pl}[(1-y)^2m^2_\mathrm{p}+ym^2_\mathrm{\pi}]}\cdot p^{n+2}+\frac{4ky(1-y)}{(1-y)^2m^2_\mathrm{p}+ym^2_\mathrm{\pi}}\cdot p -1=0.
\end{equation}
that is
\begin{equation}\label{A3}
    \begin{cases}
    G(p, y; \eta_\mathrm{p,n}, \eta_\mathrm{\pi,n}):=\alpha_n(y)\cdot p^{n+2}+\beta(y)\cdot p-1=0;\\
    \alpha_n(y; \eta_\mathrm{p,n}, \eta_\mathrm{\pi,n})=\frac{y(1-y)[\eta_\mathrm{\pi,n}(1-y)^{n+1}-\eta_\mathrm{p,n}(1-y^{n+1})]}{E^n_\mathrm{Pl}[(1-y)^2m^2_\mathrm{p}+ym^2_\mathrm{\pi}]};\\
    \beta(y)=\frac{4ky(1-y)}{(1-y)^2m^2_\mathrm{p}+ym^2_\mathrm{\pi}}.\\
    \end{cases}
\end{equation}
For a fixed energy-momentum distribution parameter $y$, $\beta(y)$ is fixed and $\beta(y)>0$; and $\alpha_n(y; \eta_\mathrm{p,n}, \eta_\mathrm{\pi,n})$ will change with different $(\eta_\mathrm{p,n}, \eta_\mathrm{\pi,n})$. 
For $p=0$, $G(0, y; \eta_\mathrm{p,n}, \eta_\mathrm{\pi,n})=-1$. 
\begin{figure}[H]
    \centering
    \includegraphics[scale=0.97]{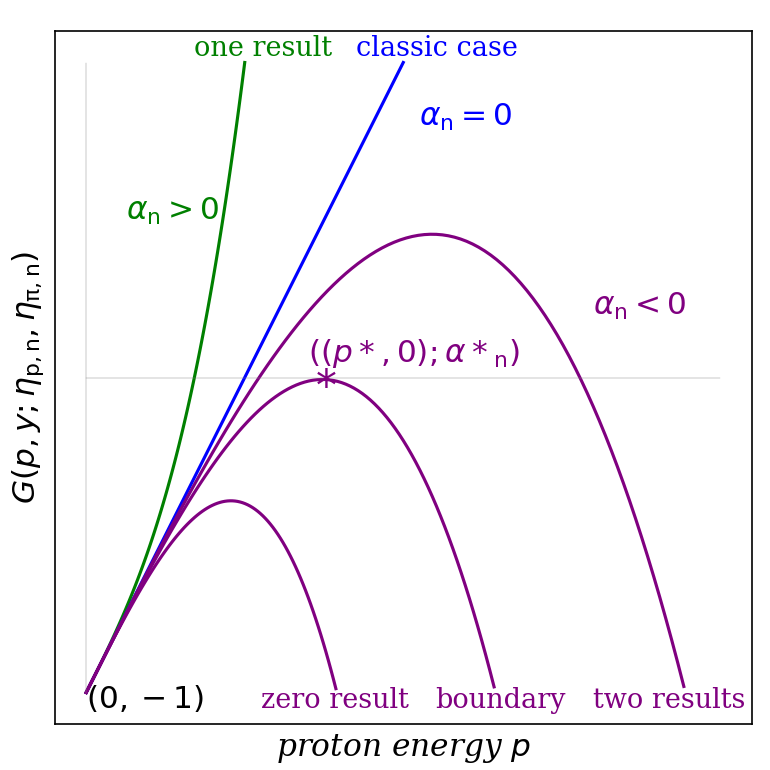}
    \caption{The trend diagram of $G(p, y; \eta_\mathrm{p,n}, \eta_\mathrm{\pi,n})$. }
    \label{fig2}
\end{figure}
For a ﬁxed $y$~($\beta(y)$ is fixed), as shown in Fig.~\ref{fig2}, there are either one, two or zero solutions to Eq.~(\ref{A3}):
\begin{itemize}
    \item 
    Case I, $\alpha_n(y; \eta_\mathrm{p,n}, \eta_\mathrm{\pi,n})\ge 0$, and there is one solution. 
    Similar to the classical situation, this solution is a candidate for the reaction threshold $p_\mathrm{th}^\mathrm{scatter}$.  
    Above this threshold, the reaction occurs and results in a drop in the cosmic-ray spectrum. 
    Unlike the classical situation, the threshold varies with the LV parameter $(\eta_\mathrm{p,n}, \eta_\mathrm{\pi,n})$. 
    When $(\eta_\mathrm{p,n}=\eta_\mathrm{\pi,n})=0$, $\alpha_n(y; \eta_\mathrm{p,n}, \eta_\mathrm{\pi,n})= 0$, which is the classical case with the solution of $p$ being the classical threshold.      
    \item 
    Case II, $\alpha_n^{*}<\alpha_n(y; \eta_\mathrm{p,n}, \eta_\mathrm{\pi,n})<0$, and there are two solutions.
    The lower one is a candidate for the lower threshold $E_\mathrm{th,low}$. 
    The upper value can correspond to a high threshold $E_\mathrm{th, high}$, that is, the highest available value for which the reaction is kinematically allowed.
    Above the low threshold $E_\mathrm{p,low}$, the reaction occurs and results in a drop in the cosmic-ray spectrum. 
    However, unlike the classical situation, there is an additional high threshold $E_\mathrm{th, high}$, above which the reaction does not occur. 
    It means cosmic rays above this high threshold can be observed. 
    From the cosmic-ray spectrum, there is a cutoff at the low threshold and a reappearance at the high threshold.
    $\alpha_n^{*}$ occurs when the top of the curve described by Eq.~(\ref{A3}) is tangent to the proton energy $p$ axis. 
    This occurs at $(p^{*}=\beta^{-1}\cdot\frac{n+2}{n+1}, \alpha_n^{*}=-\beta^{n+2}\cdot\frac{(n+1)^{(n+1)}}{(n+2)^{(n+2)}})$.
    When $\alpha_n=\alpha_n^{*}$, the tangent represents the boundary  whether Eq.~(\ref{A3}) has solutions or not.
    \item 
    Case III, $\alpha_n(y; \eta_\mathrm{p,n}, \eta_\mathrm{\pi,n})<\alpha_n^{*}$, and there is no solution. 
    In this situation, the collision reaction does not occur.
\end{itemize}

\begin{figure}[H]
    \centering
    \includegraphics[scale=1]{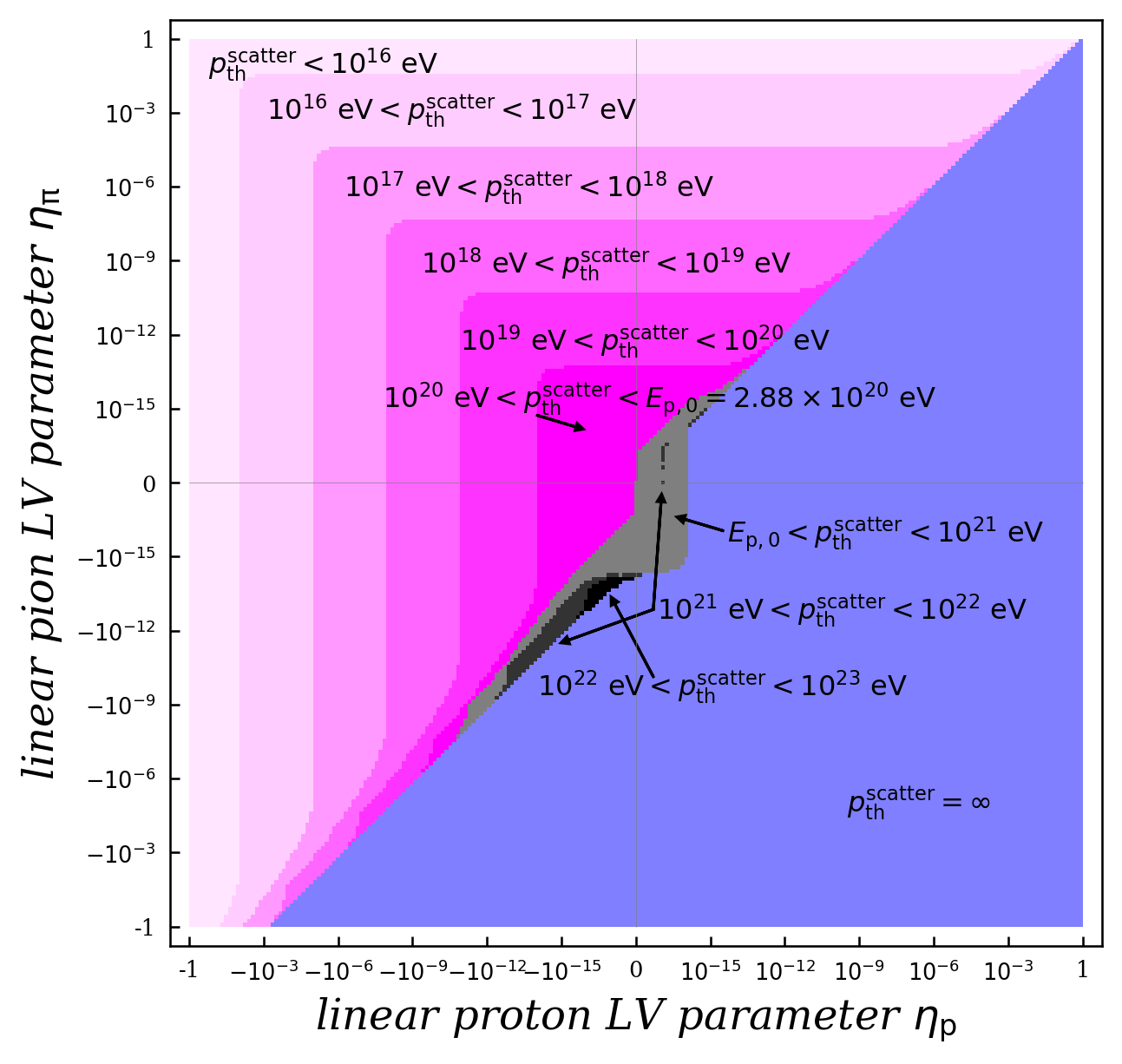}
    \caption{The minimum proton thresholds in the proton-pion LV parameter plane. 
    In this plane, every point has a fixed $(\eta_\mathrm{p,n}, \eta_\mathrm{\pi,n})$. 
    The color of each point represents the corresponding threshold value. 
    In the origin point, $(\eta_\mathrm{p,n}, \eta_\mathrm{\pi,n})=(0,0)$, and the corresponding threshold is $E_\mathrm{th}=2.88\times10^{20}~\mathrm{eV}$. 
    The red represents the threshold being less than the classical value; deeper colors show larger values. 
    The black represents the threshold being larger than the classical value; deeper colors show larger values.
    The blue represents there are no results corresponding to Case III.}
    \label{fig3}
\end{figure}

From the above discussion, we find that when $\alpha_n(y; \eta_\mathrm{p,n}, \eta_\mathrm{\pi,n})>\alpha_n^{*}$, Eq.~(\ref{A3}) has solutions, and the reaction can occur.
The threshold is the minimum momentum value $p_\mathrm{min}$ of Eq.~(\ref{A3}), for different values of the LV parameters $(\eta_\mathrm{p,n}, \eta_\mathrm{\pi,n})$ and the energy-momentum distribution parameter $y$.
We solve this problem numerically and show the results in Fig.~(\ref{fig3}).
In the plane of proton-pion LV parameter, every point has a fixed $(\eta_\mathrm{p,n}, \eta_\mathrm{\pi,n})$. 
Changing the energy-momentum distribution parameter $y$, we find the momentum minimum value $p_\mathrm{min}$ of Eq.~(\ref{A3}) at every point in the plane of the LV parameter of the proton-pion LV.\\

Fig.~\ref{fig3} shows the minimum proton thresholds at each fixed LV parameter point $(\eta_\mathrm{p,n}, \eta_\mathrm{\pi,n})$. 
The color of each point represents the corresponding threshold value.  
In the origin point, $(\eta_\mathrm{p,n}, \eta_\mathrm{\pi,n})=(0,0)$, and the corresponding threshold is $E_\mathrm{th}=2.88\times10^{20}~\mathrm{eV}$. 
The red represents the threshold being less than the classical value, where deeper colors show larger values. 
The black represents the threshold being larger than the classical value; again, deeper colors show larger values.
The blue represents that there are no results that correspond to Case III.\\

\begin{figure}[H]
    \centering
    \includegraphics[scale=1]{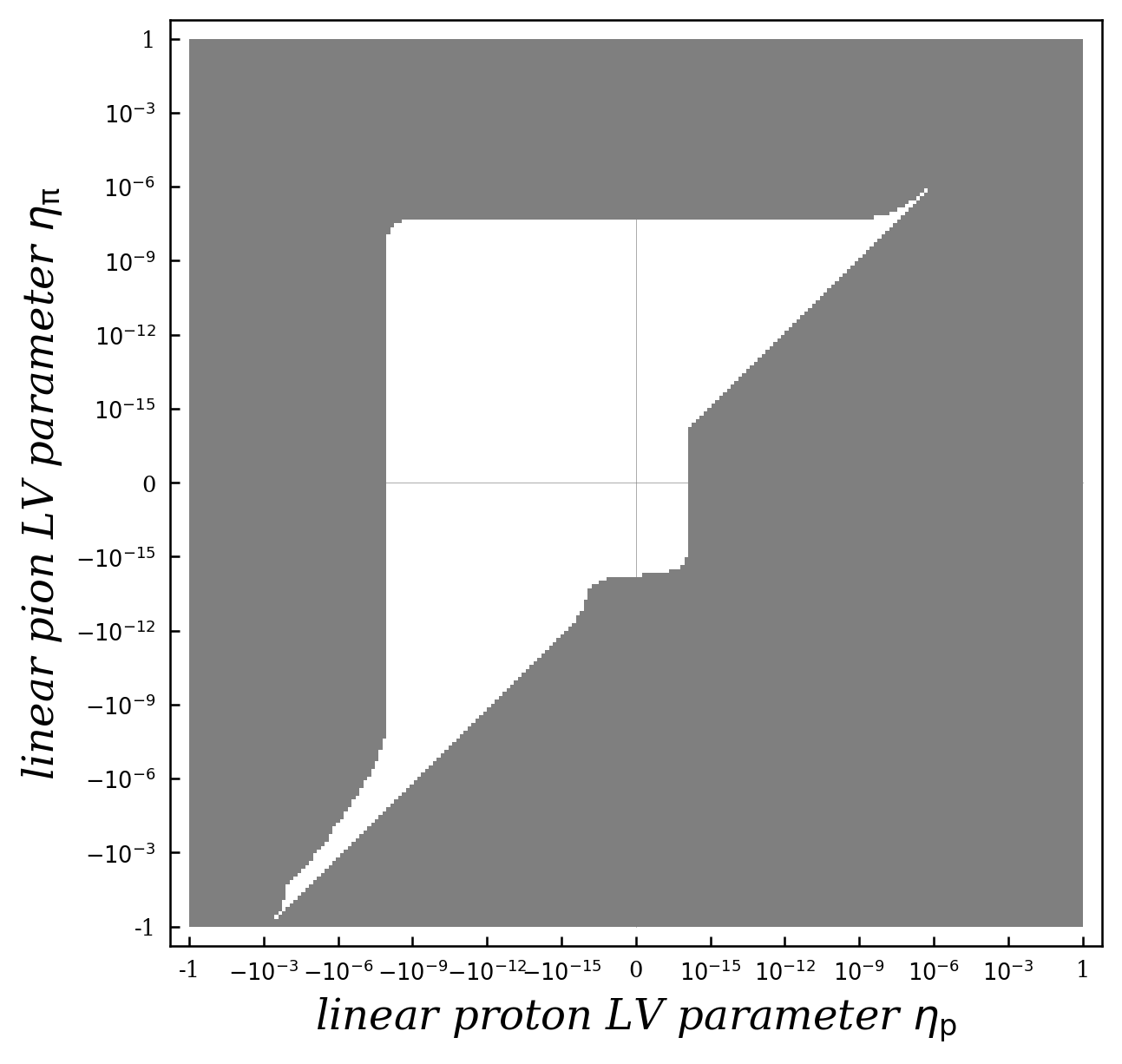}
    \caption{The constrains on the proton-pion LV parameter plane where the threshold is constrained within $10^{18}~\mathrm{eV}\sim10^{22}~\mathrm{eV}$.}
    \label{fig4}
\end{figure}

From Fig.~\ref{fig3}, we can get the constraints on the proton-pion LV parameter plane from the observations.
The existence of a threshold means that above the threshold, the reaction occurs and results in a drop in the cosmic-ray spectrum. 
The existence of GZK structure in the cosmic-ray spectrum implies strict constraints on the blue region, where the collision reaction does not occur.
In red and black regions, the minimum proton thresholds exist, so these LV parameters are all compatible with the existence of GZK structure in the observed cosmic-ray spectrum.
The constraints on the red and black regions need more exact observations on GZK structure in the cosmic-ray spectrum.
For example, if more accurate data on the GZK position are observed and the reaction threshold is constrained within $10^{18}~\mathrm{eV}\sim10^{22}~\mathrm{eV}$, then the corresponding region is strictly constrained. 
As shown in Fig.~\ref{fig4}, when the threshold is constrained within $10^{18}~\mathrm{eV}\sim10^{22}~\mathrm{eV}$, the black region is strictly constrained, and the white region is the allowed space for some theories that allow for some specific proton LV effects.\\

It is important to note that accurately determining the precise position of the GZK threshold in experiments is a challenging task that necessitates a substantial accumulation of data. 
Furthermore, the exact location of the GZK threshold does not directly correspond to the proton reaction threshold, as it must account for various factors, including the properties of the CMB, the composition of UHECRs, the origins of protons, and propagation effects.
The constraints represented in the red and black regions of Fig.~\ref{fig3} should be interpreted with caution, as their threshold behavior resembles that of classical cases, differing primarily by a change in the cutoff energy value. 
This implies that the LV parameters for protons in these regions have not been definitively excluded, leaving room for theories that allow for specific proton LV effects.
In Fig.~\ref{fig4}, we consider an ideal scenario in which the proton reaction threshold is accurately determined. 
In the blue region of Fig.~\ref{fig3}, which does not support the GZK structure, the existence of the GZK threshold imposes stringent constraints. 
Nevertheless, there remains a strong requirement for the sources of these cosmic rays: astrophysical sources must be capable of accelerating protons to the corresponding energy range.\\

The boundary of blue region in Fig.~\ref{fig3} is the boundary whether Eq.~(\ref{A3}) has solutions or not, that is, the tangent $\alpha_n=\alpha_n^{*}$.
This boundary is also the boundary of strictly constraining the LV parameter plane from the observations.
From the boundary, we get
\begin{equation}\label{A4}
    \eta_\mathrm{\pi,n}(1-y)^{n+1}-\eta_\mathrm{p,n}(1-y^{n+1})=-E^n_\mathrm{Pl}\cdot\frac{(n+1)^{n+1}}{(n+2)^{n+2}}\cdot\frac{(4k)^{n+2}y^{n+1}(1-y)^{n+1}}{[(1-y)^2m^2_\mathrm{p}+ym^2_\mathrm{\pi}]^{n+1}}.
\end{equation}
If we do not consider the pion LV effect, that is, $\eta_\mathrm{\pi,n}=0$, we get a strict constraint on the proton LV parameter $\eta_\mathrm{p,n}$
\begin{equation}\label{A5}
    \eta_\mathrm{p,n}<E^n_\mathrm{Pl}\cdot\frac{(n+1)^{n+1}}{(n+2)^{n+2}}\cdot\frac{(4k)^{n+2}y^{n+1}(1-y)^{n+1}}{[(1-y)^2m^2_\mathrm{p}+ym^2_\mathrm{\pi}]^{n+1}(1-y^{n+1})}.
\end{equation}
For $n=1$, the constraint is $\eta_\mathrm{p}<E_\mathrm{Pl}\cdot\frac{(2)^{2}}{(3)^{3}}\cdot\frac{(4k)^{3}y^{2}(1-y)^{2}}{[(1-y)^2m^2_\mathrm{p}+ym^2_\mathrm{\pi}]^{2}(1-y^{2})}$.
If we simply consider the CMB characteristic energy $\omega_0\equiv kT=2.35\times10^{-4}~\mathrm{eV}$ as the photon reaction energy, we get a strict constraint on the proton subluminal LV effect of $\eta_\mathrm{p}<1.06\times10^{-16}$~(this constraint is taken at $y=0.92$). 
If we do not consider the proton LV effect, that is, $\eta_\mathrm{p,n}=0$, we get a strict constraint on the pion LV parameter $\eta_\mathrm{p,n}$ of
\begin{equation}\label{A6}
    \eta_\mathrm{\pi,n}>-E^n_\mathrm{Pl}\cdot\frac{(n+1)^{n+1}}{(n+2)^{n+2}}\cdot\frac{(4k)^{n+2}y^{n+1}}{[(1-y)^2m^2_\mathrm{p}+ym^2_\mathrm{\pi}]^{n+1}}.
\end{equation}
Taking into account the characteristic energy of CMB $\omega_0\equiv kT=2.35\times10^{-4}~\mathrm{eV}$, we get a strict constraint on the LV effect of pion of $\eta_\mathrm{\pi}>-4.52\times10^{-15}$~(this constraint is taken at $y=1$).\\

\section{Discussion}

From the aforementioned research, we derived stringent constraints on the proton LV effect based on astroparticle phenomena. 
Specifically, for the case where $n=1$, the bilateral range for the constraint is expressed as follows
$$-\frac{m^2_\mathrm{p}E_\mathrm{Pl}}{2E_\mathrm{max}^3}<\eta_\mathrm{p}<E_\mathrm{Pl}\cdot\frac{(2)^{2}}{(3)^{3}}\cdot\frac{(4k)^{3}y^{2}(1-y)^{2}}{[(1-y)^2m^2_\mathrm{p}+ym^2_\mathrm{\pi}]^{2}(1-y^{2})}.$$
The superluminal constraint, $\eta_\mathrm{p}>-\frac{m^2_\mathrm{p}E_\mathrm{Pl}}{2E_\mathrm{max}^3}$, is derived from considerations of proton decay without accounting for photon LV effects. 
In contrast, the subluminal constraint, $\eta_\mathrm{p}<E_\mathrm{Pl}\cdot\frac{(2)^{2}}{(3)^{3}}\cdot\frac{(4k)^{3}y^{2}(1-y)^{2}}{[(1-y)^2m^2_\mathrm{p}+ym^2_\mathrm{\pi}]^{2}(1-y^{2})}$, arises from discussions of proton-photon interactions, assuming no pion LV effects.
It is important to note that we do not consider the LV effects of the final-state photons and pions when deriving these bilateral constraints for protons. 
The rationale for this assumption is that LV effects become significant only at sufficiently high particle energies. 
Given that the energy of the final-state photons and pions following the reaction is much lower than that of the initial-state protons, it is reasonable to neglect the LV effects of the final-state photons and pions when determining the LV constraints on protons. 
This simplification allows us to focus on the dominant contributions to the constraints while maintaining the robustness of our results.\\

The bilateral constraint on proton LV is significantly influenced by cosmic-ray observations. 
High-energy proton events detected via cosmic rays impose stringent constraints on the superluminal LV effect. 
For example, considering the highest-energy event recorded at $2.44~\mathrm{EeV}$ by the Telescope Array experiment~\cite{P88-TelescopeArray-2023-An}, we derive a strict superluminal constraint of $\eta_\mathrm{p}>-3.69\times10^{-16}$. 
This constraint relies on the assumption that this high-energy event is indeed a proton, which can be utilized to limit the phenomenon of proton decay.
Considering the cosmic-ray event energy errors, $2.44\pm0.29~(\rm{stats.})^{+0.51}_{-0.71}~(\rm{syst.})\times10^{20}~\mathrm{eV}$~\cite{P88-TelescopeArray-2023-An}, we get the corresponding error of the constraints on the superluminal LV effect: $\eta_\mathrm{p}>-3.69^{+1.06}_{-1.71}~(\rm{stats.})^{+1.60}_{-6.67}~(\rm{syst.})\times10^{-16}$. We see that the measurement uncertainties of the energies of cosmic-ray events directly lead to the uncertainties in the constraints. 
The energy uncertainties for this event in the Telescope Array experiment arise from two main sources: (1) the statistical uncertainty in the reconstructed energy, which has an energy resolution of $29~\mathrm{EeV}$ for this event~\cite{P106-TelescopeArray-2012-the}, and (2) systematic uncertainty from the surface detector array in the Telescope Array experiment, which is $21\%$~\cite{P107-TelescopeArray-2015-the}, along with an additional $10\%$ systematic uncertainty in the lower energy direction, arising from the unknown primary, as estimated from simulations~\cite{P88-TelescopeArray-2023-An}.\\

The existence of the GZK structure establishes a clear boundary for subluminal proton LV effects. 
By considering the characteristic energy of the CMB, defined as $\omega_0\equiv kT=2.35\times10^{-4}~\mathrm{eV}$, we obtain a stringent constraint on the subluminal proton LV effect: $\eta_\mathrm{p}<1.06\times10^{-16}$. 
This constraint is predicated on the important assumption that the pion production process dominates the structural changes observed in the cosmic-ray GZK region.
This assumption entails several critical factors: first, astrophysical sources are capable of accelerating protons beyond the GZK energy range; second, protons constitute the dominant component in the GZK region; and third, structural changes in this region primarily arise from the propagation effects resulting from interactions with the CMB.
The acceleration processes occurring at the sources are fundamental in determining the initial composition of cosmic-ray components. 
Generally, the efficiency of acceleration is proportional to the nuclear charge, suggesting that heavier elements are boosted to higher energies compared to lighter ones, although all elements ultimately face a maximum energy limit.
Recent experiments, such as HiRes~\cite{P50-HiRes-2009-indications}, the Pierre Auger Observatory~\cite{P51-PierreAuger-2010-measurement, P52-PierreAuger-2014-depth}, and the Telescope Array~\cite{P53-TelescopeArray-2018-depth}, have provided valuable measurements of cosmic-ray composition. 
However, there remains no consensus regarding the exact composition in the GZK region: while the results from HiRes and the Telescope Array are consistent with a light primary composition predominantly composed of protons, the data from the Auger Observatory suggest a heavier composition. 
This discrepancy highlights the complexity of cosmic-ray composition studies and underscores the need for further research to clarify these findings.
The uncertainties in the constraints on the subluminal LV effect are directly influenced by the uncertainties in the CMB experimental values. According to Eq.~(\ref{A5}), the uncertainties propagate to the third power, that is, if the CMB energy error is $x\%$, the resulting constraint error is $[(1+x\%)^3-1]$.
We do not provide a quantitative description of the constraint uncertainty here, as the primary source of uncertainty arises from the randomness in the interaction of CMB photons with cosmic-ray events. A detailed quantitative analysis of the constraint uncertainty would require a more precise calculation of the proton's mean free path in the CMB background.\\

Our study presents a bilateral constraint for protons, defined by the range $-3.69\times10^{-16}<\eta_\mathrm{p}<1.06\times10^{-16}$, which encompasses both subluminal and superluminal scenarios. 
However, more rigorous data and experiments are necessary to substantiate these findings.
The Pierre Auger Observatory has established constraints on LV in both the electromagnetic and hadronic sectors by comparing observational data with LV simulations of the energy spectrum, cosmic-ray composition, and upper limits on photon flux~\cite{P55-Lang-2020-Testing, P57-PierreAuger-2021-testing}. 
Specifically, the Observatory examines the propagation of UHECR nuclei, taking into account LV effects in photopion production and photodisintegration to assess LV in the hadronic sector.
The methodology employed by the Pierre Auger Observatory is statistically rigorous, incorporating systematic error analysis for both observational data and propagation simulations. 
This approach ensures that their constraints on LV are not only statistically significant but also reproducible.
In contrast, while the Pierre Auger Observatory provides a superluminal LV constraint in the hadronic sector with enhanced robustness and rigor, our constraint is bilateral, encompassing both subluminal and superluminal scenarios. This distinction highlights the complementary nature of our findings in the broader context of LV research.\\

Further theoretical and observational advancements are essential for investigating the proton LV effect.
From a theoretical perspective, it would be beneficial to conduct a more in-depth analysis of the proton and photon reaction channels, such as $p+\gamma\to\Delta(1232)$. 
Additionally, a thorough exploration of the interactions between CMB photons and protons is crucial. 
In classical scenarios, the mean free path of protons through CMB photons decreases exponentially with energy, dropping to a few megaparsecs above the GZK cutoff. 
In the context of LV, detailed information about the reaction thresholds and the mean free path of protons would be particularly valuable.
On the observational side, a deeper understanding of the sources, composition, and energies of UHECRs is required. 
Accurate future measurements of UHECRs will significantly enhance our research into LV. 
For instance, the upcoming upgrade of the Pierre Auger Observatory, known as AugerPrime~\cite{P99-Martello-2017-the,P100-Cataldi-2021-the}, is expected to improve our understanding of cosmic-ray composition, which will be crucial for studies of LV.
Research on LV has the potential to provide valuable insights into quantum gravity and grand unified theories, particularly those that accommodate specific LV modifications.\\

Our discussion is framed within a specific context where the conservation of energy and momentum remains intact, and LV effects are introduced solely through modifications to the dispersion relations. However, the interplay between conservation laws and LV effects is more intricate than this approach suggests. Energy and momentum conservation are fundamentally linked to Lorentz symmetry, and any modifications to this symmetry could introduce new effects that may influence these conservation laws.
For instance, in the framework of double special relativity (DSR)~\cite{C24-amelino-2001-relativity,C25-amelino-2010-DSR}, the relationship between active variables in different coordinate systems is governed not by the conventional Lorentz transformation, but by a deformed version of it. In this context, the conservation laws are also modified: they do not adhere to the traditional form but instead are expressed through a deformation that reflects the altered symmetry.
While conservation laws remain foundational to all physical interactions and apply universally, certain LV scenarios can lead to deviations from standard energy and momentum conservation due to changes in the dispersion relations. In our study, we uphold the principle of energy and momentum conservation, introducing LV effects exclusively through modifications to these dispersion relations.
This approach offers several advantages: it allows for a clearer quantification of the impact of LV effects, avoids potential conflicts with more complex theoretical models, and provides a straightforward framework for discussing LV constraints efficiently. By focusing on this simplified model, we aim to elucidate the implications of LV without the complications that arise from more elaborate formulations.\\

When considering the dispersion relations of composite particles such as protons, it is essential to prioritize the principles of overall energy and momentum conservation. 
These fundamental principles are the cornerstones of any physical interaction and apply universally, regardless of the internal structure of the particle involved.
From a first-principles perspective, the conservation of energy and momentum remains paramount in discussions of LV effects.
While the composite nature of protons—comprising quarks and gluons—can introduce additional complexities, it does not fundamentally alter the underlying conservation laws that govern particle interactions. 
The dispersion relations derived from these conservation laws can still be valid and applicable, even when considering composite particles.
In this context, the compositeness of the proton may not play a significant role in the initial analysis of LV effects. 
The modified dispersion relations can be formulated based on the conservation laws without requiring detailed knowledge of the proton's internal structure. 
Thus, while it is crucial to acknowledge the proton's composite nature for a more nuanced understanding of its behavior, the first consideration of LV effects can be adequately addressed through the lens of energy and momentum conservation.
However, it is also important to recognize that as we delve deeper into the implications of LV and its manifestations, the composite structure of protons may become relevant. 
For example, when considering specific interactions, decay processes, or high-energy collisions, the internal dynamics of the proton could influence the outcomes~\cite{P102-Hwang-2005-neutrons}. 
Therefore, while the initial discussions of LV can proceed without a detailed consideration of compositeness, subsequent analyses should indeed take into account the complexities introduced by the proton's quark-gluon structure to enhance the robustness and accuracy of the model.\\

We need to address the implications of proton stability for LV searches. 
Proton stability is indeed a well-explored area of research, with significant constraints established through both collider experiments and cosmological observations. 
For instance, the Super-Kamiokande experiment has set a lower limit on the proton lifetime at $1.6\times10^{34}~\mathrm{yr}$  for specific decay modes~\cite{P101-Super-Kamiokande-2016-search}.
In discussing proton stability and its implications for LV searches, it is essential to recognize that fundamental principles such as energy and momentum conservation, along with baryon and lepton number conservation, serve as the foundational pillars of particle physics. 
These principles must be satisfied in any decay process, including those involving protons.
When we consider LV effects, the introduction of modified dispersion relations does not inherently violate these basic conservation laws. 
Instead, LV typically modifies the relationships between energy and momentum, which may lead to new dynamics in particle interactions but does not negate the fundamental conservation principles that govern them. 
As such, while LV may introduce novel phenomena, it does not necessarily imply that protons will exhibit behavior inconsistent with their established stability.
Given this context, we argue that the existing constraints on proton stability, such as the long proton lifetime established by experiments like Super-Kamiokande, are unlikely to impose significantly stronger constraints on LV effects in protons. 
The stability of protons, as demonstrated by these experiments, indicates that any LV-induced decay processes would need to operate within the framework of the existing conservation laws. 
Since these laws remain intact in the presence of LV, we do not expect that the stability of protons will lead to more stringent constraints on LV beyond what has already been established.
Moreover, any potential LV effects would need to manifest in a manner that is consistent with the observed stability of protons. 
For instance, if LV were to alter decay rates, it would still have to conform to the bounds set by current experimental results. 
This suggests that while LV can provide new avenues for exploration, the constraints derived from proton stability may not necessarily tighten significantly, as the modifications introduced by LV do not inherently conflict with the fundamental conservation laws.\\

\section{Conclusion}

We have obtained stringent constraints on the proton LV effect, specifically for the case of $n=1$. 
The range for the constraint is given bilaterally by 
$$-\frac{m^2_\mathrm{p}E_\mathrm{Pl}}{2E_\mathrm{max}^3}<\eta_\mathrm{p}<E_\mathrm{Pl}\cdot\frac{(2)^{2}}{(3)^{3}}\cdot\frac{(4k)^{3}y^{2}(1-y)^{2}}{[(1-y)^2m^2_\mathrm{p}+ym^2_\mathrm{\pi}]^{2}(1-y^{2})}.$$
These constraints are stringent for both subluminal and superluminal LV effects. 
The superluminal constraint $\eta_\mathrm{p}>-\frac{m^2_\mathrm{p}E_\mathrm{Pl}}{2E_\mathrm{max}^3}$ is derived from discussions on proton decay and high-energy proton events observed in cosmic rays. 
For instance, considering the highest-energy proton event at $2.44~\mathrm{EeV}$ from the Telescope Array experiment~\cite{P88-TelescopeArray-2023-An}, we obtain a strict superluminal constraint of $\eta_\mathrm{p}>-3.69\times10^{-16}$.
On the other hand, the subluminal constraint $\eta_\mathrm{p}<E_\mathrm{Pl}\cdot\frac{(2)^{2}}{(3)^{3}}\cdot\frac{(4k)^{3}y^{2}(1-y)^{2}}{[(1-y)^2m^2_\mathrm{p}+ym^2_\mathrm{\pi}]^{2}(1-y^{2})}$ is derived from discussions on the interaction of UHECRs with CMB radiation. 
The presence of the GZK structure provides a strict boundary for the peculiar phenomena arising from subluminal proton LV effects.
Considering the CMB characteristic energy $\omega_0\equiv kT=2.35\times10^{-4}~\mathrm{eV}$ as the photon reaction energy, we obtain a stringent constraint on the subluminal proton LV effect of $\eta_\mathrm{p}<1.06\times10^{-16}$. 
Therefore we obtain the bilateral constraints on proton LV effects with $-3.69\times10^{-16}<\eta_\mathrm{p}<1.06\times10^{-16}$.
\\

This work is supported by National Natural Science Foundation of China under grants No.~12335006 and No.~12075003.\\

\bibliographystyle{unsrt}

\bibliography{r.bib}{}
\end{document}